\documentstyle[epsfig,emulateapj]{article}

\def\ltsima{$\; \buildrel < \over \sim \;$}
\def\lsim{\lower.5ex\hbox{\ltsima}}
\def\gtsima{$\; \buildrel > \over \sim \;$}
\def\gsim{\lower.5ex\hbox{\gtsima}}
\def\bx#1{\leavevmode\thinspace\hbox{\vrule\vtop{\vbox{\hrule\kern1pt
        \hbox{\vphantom{\tt/}\thinspace{\bf#1}\thinspace}}
      \kern1pt\hrule}\vrule}\thinspace}

\def\thetag{\vec \theta }
\begin{document}

\title{Statistics of Dark Matter Halos from Gravitational Lensing}
\author{Bhuvnesh Jain$^1$ and Ludovic Van Waerbeke$^2$} 
\affil{{}$^1$Dept. of Physics, Johns Hopkins University, Baltimore, 
MD 21218\\ {}$^2$Canadian Institute for Theoretical Astrophysics, 
60 St George St, Toronto, M5S 3H8, Canada}

\begin{abstract}
We present a new approach to measure the mass function of
dark matter halos and to discriminate models with differing 
values of $\Omega$ through weak gravitational lensing. 
We measure the distribution of peaks from simulated lensing surveys 
and show that the lensing signal due to dark matter halos can 
be detected for a wide range of peak heights. 
Even when the signal-to-noise is well below the 
limit for detection of individual halos, 
projected halo statistics can be constrained
for halo masses spanning galactic to cluster halos. 
The use of peak statistics relies on an
analytical model of the noise due to the intrinsic 
ellipticities of source galaxies. The noise model has been shown 
to accurately describe simulated data for a variety of input ellipticity 
distributions. 
We show that the measured peak distribution has distinct signatures of
gravitational lensing, and its non-Gaussian shape can be used to 
distinguish models with different values of $\Omega$. The
use of peak statistics is complementary to the measurement of field 
statistics, such as the ellipticity correlation function, 
and possibly not susceptible to the same systematic errors. 
\end{abstract}


\section{Introduction}

In the coming years, gravitational lensing is likely to become
an effective tool for mapping large-scale structure in the 
universe. Over the past decade several measurements of weak lensing
by galaxy clusters have been made. Mass reconstruction techniques 
are now being applied to wide field lensing surveys in blank fields 
that will probe the dark 
matter distribution over angular scales of order $1' - 1^\circ$. 
Wide field lensing observations have already detected filaments 
and dark halos that were not visible by their light distribution
(Kaiser at al 1998; Erben et al 1999; Tyson et al. 1999).

Statistical properties of the clustering of dark matter 
can be probed from lensing data by computing shear
correlations over blank fields with area of order 10 square 
degrees (Blandford
et al 1991; Miralda-Escud\'e 1991; Kaiser 1992; Bernardeau et al
1997; Jain \& Seljak 1997; Kaiser 1998; Stebbins 1996; 
Schneider et al.\ 1998). 
An alternative approach is to focus on the statistics of dark matter halos, 
identified through their lensing strength, using measures such as
the aperture mass (Schneider 1996; Kruse \& Schneider 1999a; Kruse 
\& Schneider 1999a; Reblinsky et al. 1999). The halo statistics approach
has been shown by the above authors to be a useful probe of the mass
function for massive, cluster sized halos; the main practical 
limitation is that only $\sim 10$ halos per square 
degree are expected to be detected with adequate signal-to-noise. 

This paper advocates a new approach to the measurement of the 
statistics of dark matter halos through lensing. By modeling the
distribution of peaks in lensing data induced by the noise due to the
intrinsic ellipticities of source galaxies,
we show that it is possible to statistically detect the
signal due to dark matter halos, even for mass scales 
below the signal-to-noise limit for the detection of individual halos. 
Section 2 describes the construction of peak statistics from simulated
data and from pure noise. Results for the peak statistics 
for a set of cosmological models are shown in Section 3. We discuss
the prospects for measuring the halo mass function and discriminating
models from realistic data in Section 4.

\begin{figure*}[t]
\vspace{6cm}
\caption{Histogram of the peak distribution in a noisy
convergence field for the open model. 
The solid line with error bars is the peak distribution measured from
the $\kappa$ map with Gaussian noise added as
discussed in the text. The dashed line is the peak distribution measured
from the $\kappa$ map reconstructed from the ellipticities. The near
coincidence of the two curves demonstrates the accuracy of the 
reconstruction scheme and of the noise model. The error 
bars are computed from 7 realizations of the signal $\kappa$ map. 
The dot-dashed line shows the peak distribution due to the noise
alone. Almost overlapping with it is the analytical model for noise peaks
discussed in the text. 
}
\includegraphics{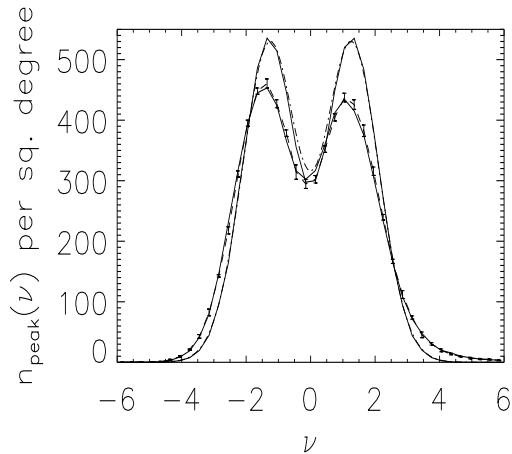}
\label{fignoise1}
\end{figure*}

\begin{figure*}[t]
\vspace{6.5cm}
\caption{Comparison of the reconstructed and input maps of $\kappa$. 
The left panel shows the input $\kappa$ maps for a field 3 degrees
on a side, and the right panel shows
the same field reconstructed from ellipticity data that includes the
noise due to the intrinsic ellipticities of galaxies. 
While the small scale peaks in
the right panel are dominated by the noise, the subsequent figures
show how their distribution contains imprints of the signal. 
}
\includegraphics{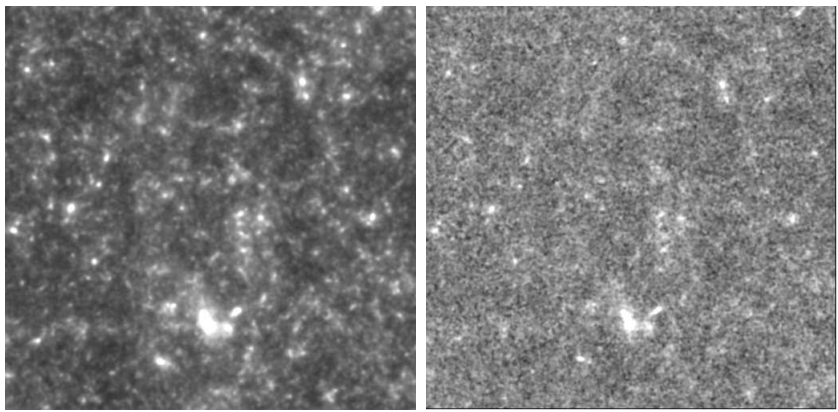}
\label{fignoise2}
\end{figure*}

\section{Peak Statistics in Simulated Data}

We use shear and convergence fields from ray tracing simulations
through the dark matter distribution of N-body simulations (Jain,
Seljak \& White 1999). 
The fields we use are about 3 degrees on a side, sampled with a grid
spacing of $0.1'$, with source galaxies taken to be at $z=1$. 
We use two cosmological models, an Einstein-de
Sitter model
and an open model with $\Omega_{\rm matter}=0.3$. 
The power spectrum corresponds to a cold dark matter 
shape parameter $\Gamma=0.21$ model. Further details of the models
and the simulations are given in Jain et al (1999). 

A simulated noisy map of the convergence, $\kappa({\vec \theta})$, is built 
by first smoothing the $\kappa$
field over scale $\theta_G$ with a Gaussian window
$W(\theta)=\exp(-{|\thetag|^2/\theta_G^2})/\pi \theta_G^2$. 
The noise due to the randomly oriented intrinsic ellipticities of source
galaxies is modeled as 
a Gaussian random field with variance, 
\begin{equation}
\sigma_{\rm noise}^2={\sigma_\epsilon^2\over 2}{1\over 2\pi\theta_G^2n_g},
\label{noise}
\end{equation}
where $\sigma_\epsilon$ is the rms amplitude of the intrinsic
ellipticity distribution and $n_g$ is the number density of source
galaxies. This Gaussian noise is added to the smoothed $\kappa$
field; the accuracy of this noise model is discussed below.
From the smoothed noisy data, peaks are 
found by identifying pixels
that have a higher/lower value of $\kappa$  than all neighboring
pixels. This corresponds to the condition that the gradient of the
field vanishes and thus includes peaks as well as troughs. 
The height of the peak $\nu=\kappa/\sigma_{\rm noise}$ is its
value in units of the noise rms in the smoothed field. 

Our choice of the noise model for the convergence field relies on
previous work. 
Van Waerbeke et al (1999) have shown that the convergence field
can be accurately reconstructed from observed ellipticity data in
the absence of systematic errors. Both the reconstruction schemes of
Kaiser \& Squires (1993) and the Maximum-Likelihood algorithm of
Bartelmann et al (1996) recover the convergence field with adequate
accuracy for fields of order a degree on a side. 
Van Waerbeke (1999) further showed that the noise properties of peaks
can be analytically described using Gaussian statistics 
(Bardeen et al 1986; Bond \& Efstathiou 1987), and the weak 
lensing approximation. 

\begin{figure*}[t]
\vspace{11cm}
\caption{Probability distribution function 
of the peaks in noise-free fields (left
panels) and noisy fields (right panels) 3 degrees on a side, with
smoothing scale $\theta_G=0.5'$ (upper panels) and $1'$ (lower panels). 
The solid lines show the open model and the dashed lines the EdS 
model. The error bars are obtained from averaging over seven
realizations. In the right panels the dot-dashed lines show the 
pdf of peaks in fields with pure noise. 
}
\includegraphics{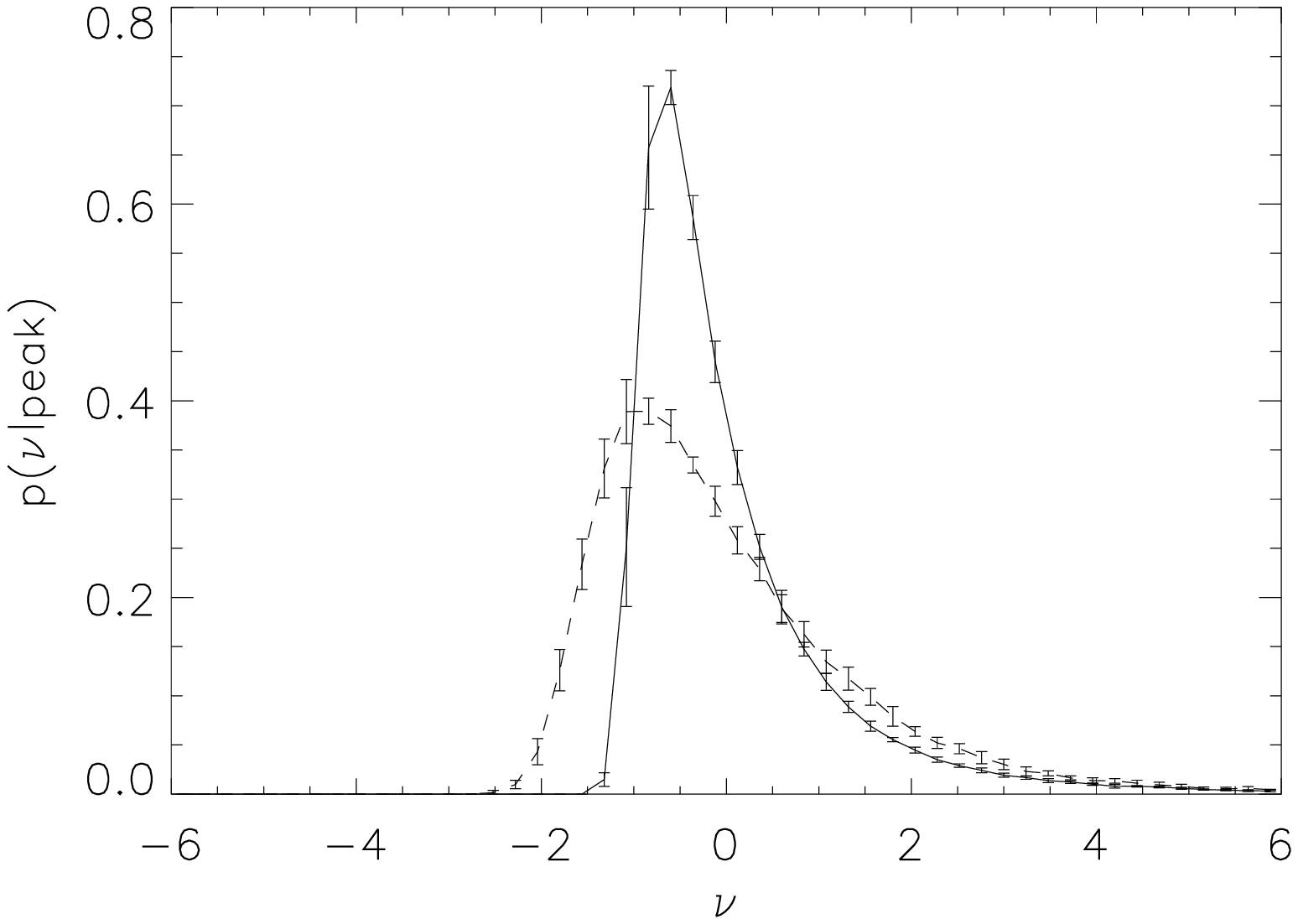}
\includegraphics{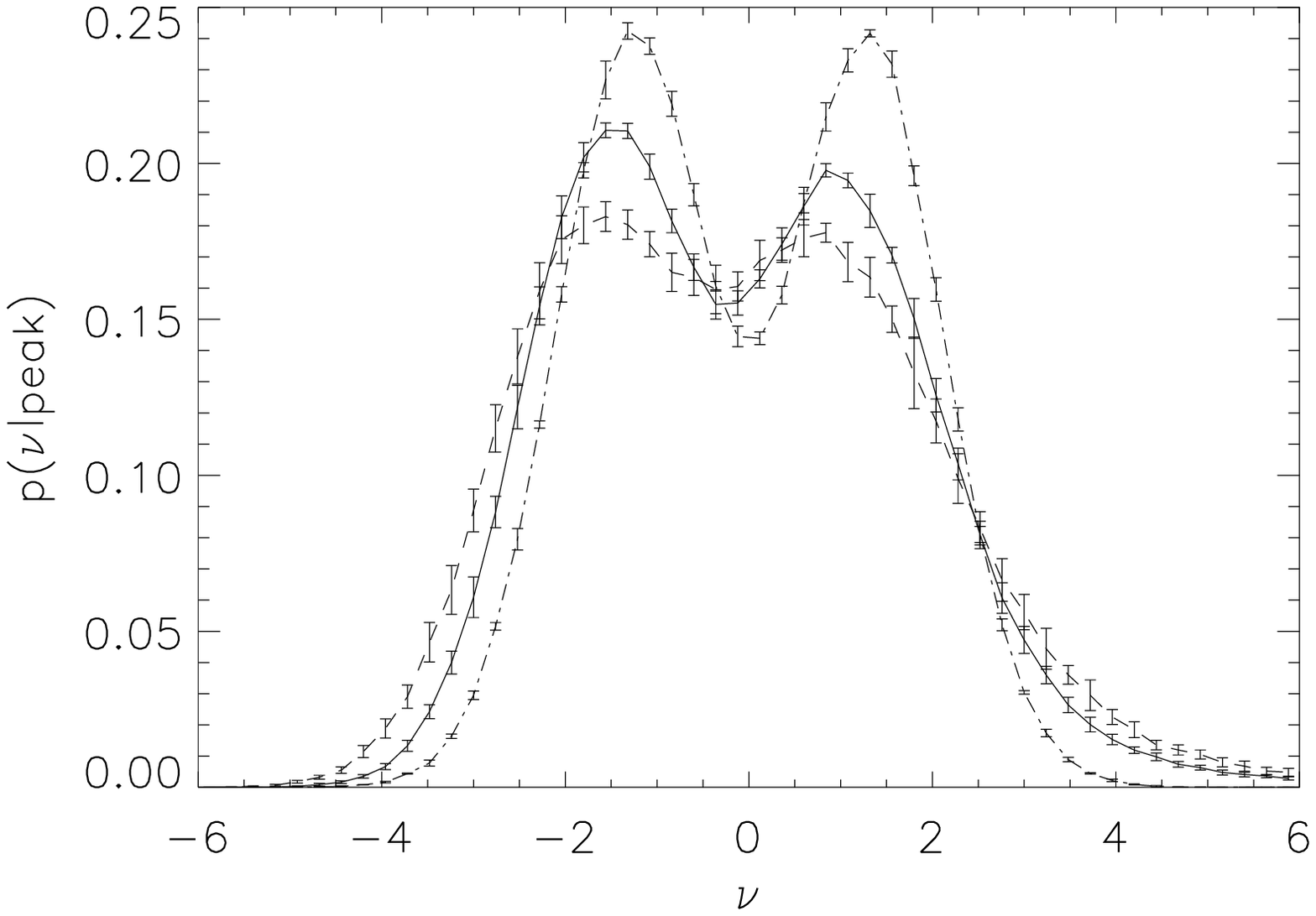}
\includegraphics{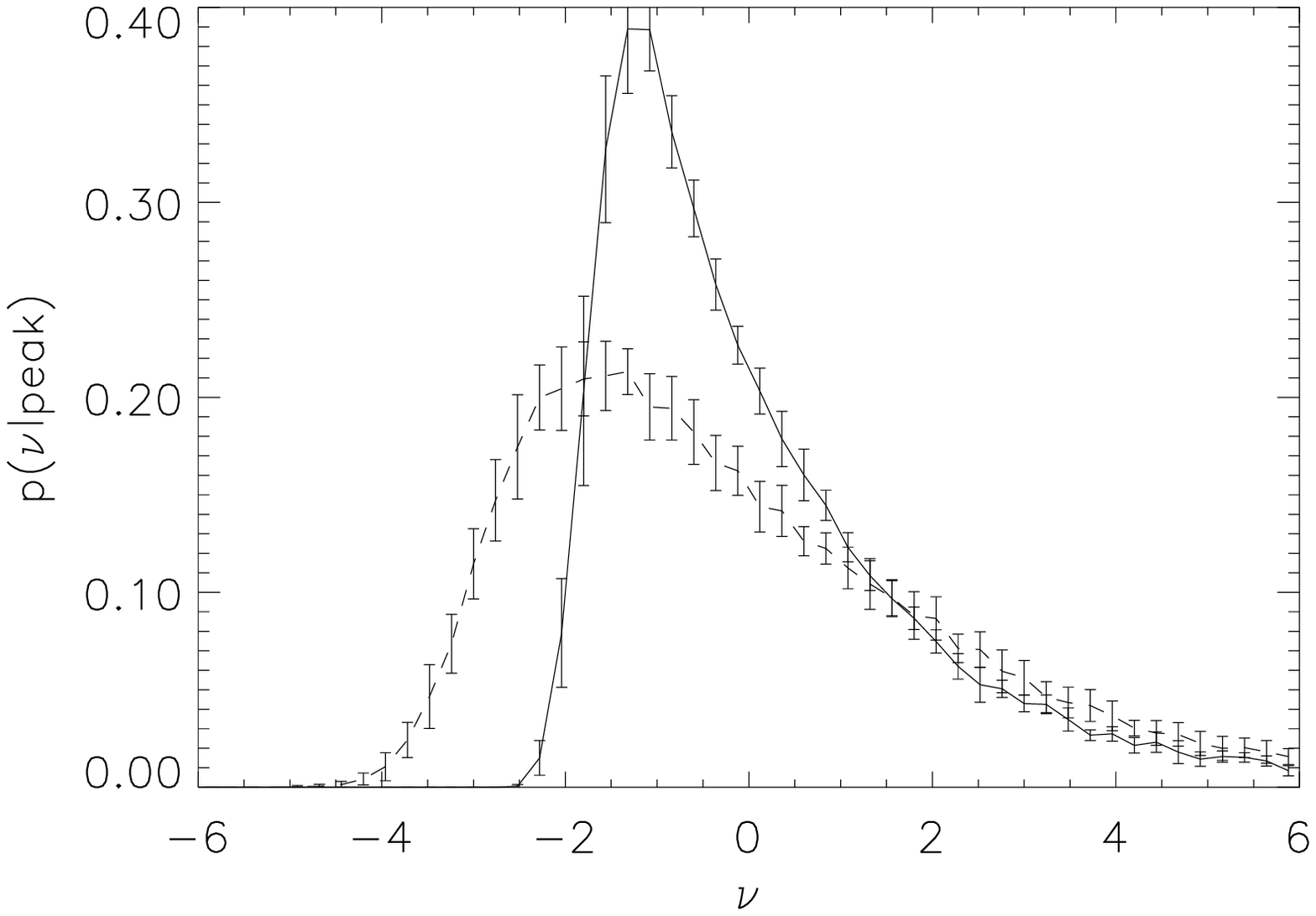}
\includegraphics{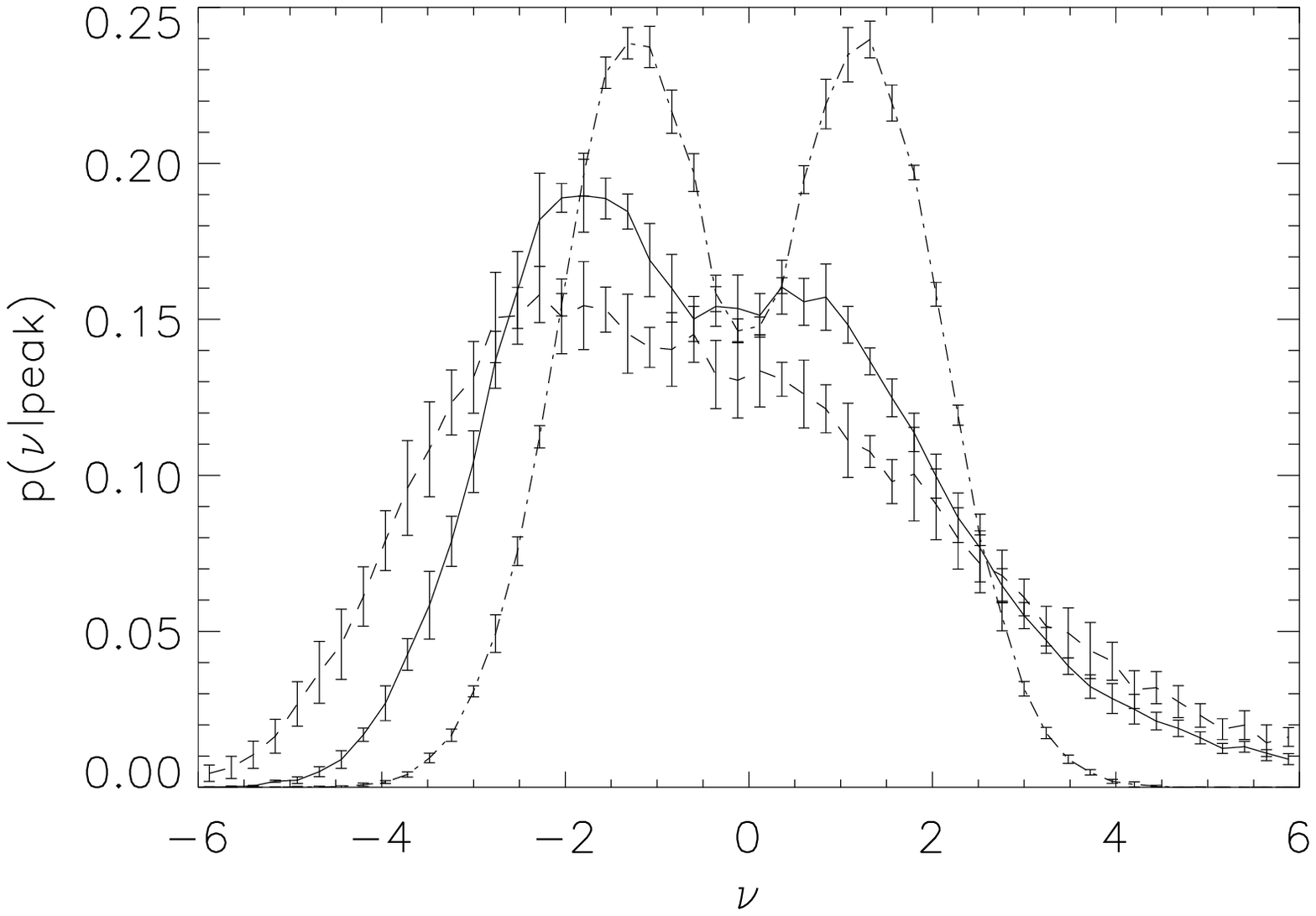}
\label{fighist}
\end{figure*}

Figure \ref{fignoise1} shows a test of the analytical model of 
Van Waerbeke (1999) for peaks due to noise, and
checks the accuracy of the peak distribution in the reconstructed $\kappa$. 
The dot-dashed curve 
shows the histogram of peaks from this analytical noise
model; the double peaked shape is due to peaks with
positive curvature and troughs with negative curvature. 
Almost overlapping with the analytical curve is the measured 
histograms of peaks in a field with pure noise. 
The distribution of peak heights
measured in maps of $\kappa$ reconstructed from noisy ellipticity data
(dashed line) is compared with the distribution
in the $\kappa$ maps of the signal plus Gaussian noise with variance
given by equation \ref{noise} (solid line). The close 
agreement of the two curves
demonstrates the accuracy of the $\kappa$ reconstruction scheme. 

The success of the reconstruction gives us
confidence in working with the convergence data and the noise model 
directly, avoiding the slow and expensive reconstruction process. 
We have also verified that using the mass aperture statistic, 
which is constructed directly from ellipticity data, leads to very
similar peak distributions. 
We have also found that a variety of distributions of the intrinsic
ellipticity (including non-Gaussian ditributions) 
produce the same Gaussian statistics for
peaks in the $\kappa$ maps; these results will be presented elsewhere. 
Figure \ref{fignoise2} shows the actual maps
of the convergence, $\kappa$, used to measure the peak distributions. 
The small amplitude peaks in the signal map are swamped by the noise, 
so there is little hope of recovering them individually from
data. However, we show below that their distribution 
is sufficiently modulated by the signal to distinguish cosmological models. 


\section{Sensitivity of Peak Statistics to $\Omega$}

\begin{figure*}[t]
\vspace{5.5cm}
\caption{Histogram of the convergence field in noise-free fields (left
panel) and noisy fields (right panel). 
The solid lines show the open model and the dashed lines the EdS 
model. The error bars are obtained from averaging over seven
realizations. The field size and other parameters are as for the upper
panels of figure \ref{fighist}.
}
\includegraphics{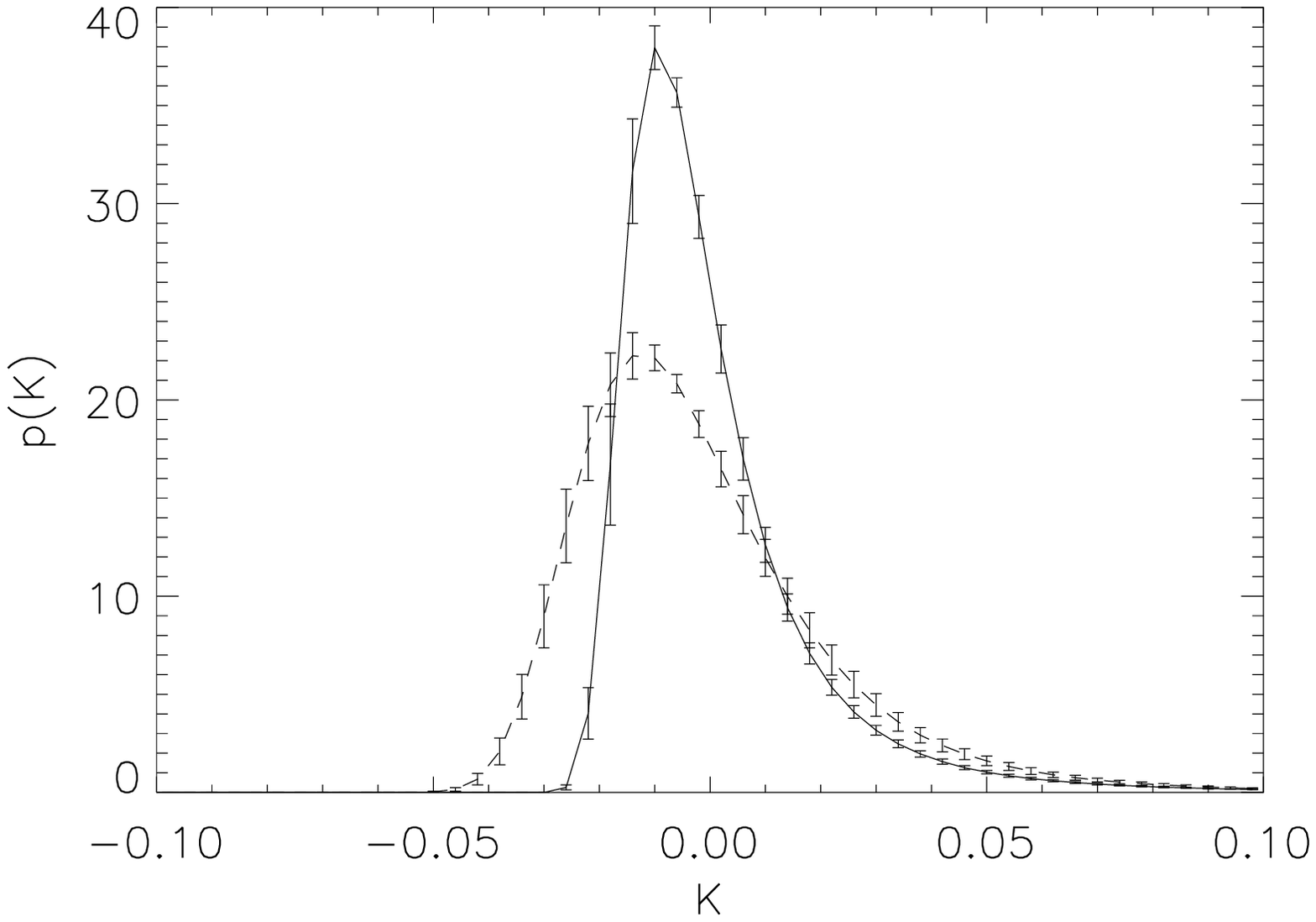}
\includegraphics{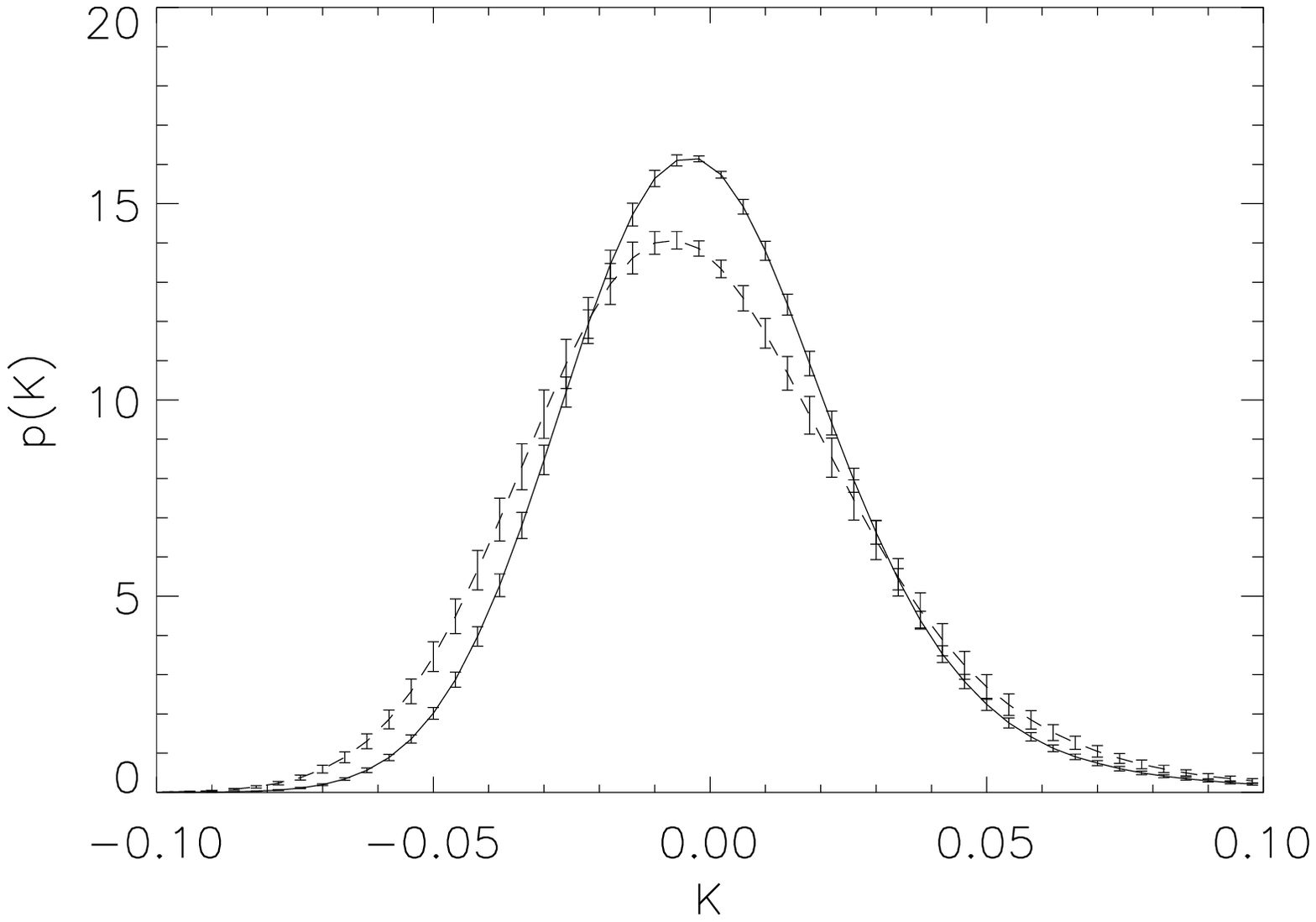}
\label{figfield1}
\end{figure*}

Figure \ref{fighist} shows the probability distribution function
(pdf) of peaks in the convergence field 
as a function of peak height for noise-free (left
panels) and noisy (right panels) fields for two different smoothing 
scales. The pdf in noise-free fields has the qualitative
characteristics due to nonlinear gravitational clustering: at negative
peak heights (underdense regions) it has a cutoff related
to the minimum $\kappa$ resulting from empty beams,
and it has a tail at positive $\nu$ due to 
collapsed halos. The cosmological models have different pdf's, just 
as they do for the pdf of $\kappa$ in the field, shown in figure
\ref{figfield1}.
For the noisy fields, the number density of
source galaxies is  30 per square arcminute, and their rms 
intrinsic ellipticity is $\sigma_\epsilon=0.2$. 
Thus the peak height $\nu=1$ corresponds to an averaged value
of $\kappa=0.02\ (0.01)$ over the smoothing radius
for the upper (lower) panels of figure \ref{fighist}. 

The right panels of figure \ref{fighist} show that in the
presence of noise, the pdf's look quite different from the noise-free
case. However the noisy pdf's still have different shapes from the 
pure noise pdf's and the cosmological models remain distinguishable.
The asymmetric double peak for the low amplitude peaks ($-2<\nu<2$)
arises due to the noise maxima and minima, but it is suppressed relative to
the pure noise case and is asymmetric due to the gravitational shear. 
The open and Einstein-de Sitter models are easily distinguishable for
these low amplitude peaks, even though almost none of the peaks
can be individually associated with dark matter halos. 
The relative number 
of very negative troughs (which are mostly noise troughs modulated 
by the fact that they are located in large voids) 
can by itself be used to
discriminate models with different values of $\Omega$, as
noted by Jain et al (1999) for the field pdf. 
As the smoothing scale is increased from
$0.5'$ to $1'$, the signal dominates
over the shape of the noise pdf, but sample variance becomes
larger as there are fewer peaks. As a result, at both smoothing
scales, the models can be distinguished at about the same level
of significance. The error bars in the right panels are not much
larger than in the pdf from the noise-free maps. The addition of
ellipticity noise broadens the peak distribution, but the error bars
are still dominated by sample variance in the signal. 
Similary, the primary effect of increasing $\sigma_\epsilon$
is to broaden the pdf while not changing the error bars by much. 

It is worth noting that previous theoretical work on halo detection
has focused on the peaks that can be individually detected with
adequate signal to noise. These would correspond to the parts of
the pdf with $\nu\gsim 4-5$, where sample variance is large. 
Clearly the bulk of the information on the mass function and in 
distinguishing models is at smaller or negative peak heights and 
can be used only statistically by modeling the noise pdf. 

\section{Discussion}

The results presented in sections 2 and 3 show that the 
peak distribution from lensing data has information on
the projected mass function of dark matter halos, and is 
sensitive to the cosmological model. 
The level of non-Gaussianity of the pdf is a powerful discrimant of
models with different values of $\Omega$. Figure \ref{fighist} shows that the
models can be distinguished from the pdf over a wide range of peak 
heights at 2 to 3-$\sigma$; by combining information at different peak
heights and smoothing scales we can obtain much higher significance. 
Further, the third and fourth moments of the peak distribution for
different smoothing scales are sensitive to the cosmological model, as
expected qualitatively from the shapes of the distribution. 
We have
also compared the peak distributions shown with a model with
non-zero cosmological constant $\Omega_\Lambda=0.7$. 
The peak distribution in the $\Lambda$-model lies in-between the
Einstein de-Sitter and open model with the same value of $\Omega_{\rm 
matter}$.

Beyond the dependence on the cosmological parameters, the peak pdf
contains information on the projected mass function over all mass scales.
It is important to test how accurately we can recover the pdf of peaks 
due to the lensing signal, and hence the projected mass function, 
from wide field lensing surveys. A straightforward approach is 
to compare the measured pdf with the predictions of a set of 
models that include the level of noise observed in the data. The
best fit model can be found by minimizing the $\chi^2$. 
We demonstrate in a forthcoming paper that the projected mass 
function and $\Omega$ can be simultaneously determined
by using the normalization and shape of the distribution (Van 
Waerbeke \& Jain 1999). Since we use information from all peak
heights, not just the high-$\sigma$ peaks that can be detected
individually, the mass function is constrained 
over mass scales ranging from galactic to cluster sized halos. 
A more ambitious approach to recover the lensing signal 
would be to de-convolve the measured peak 
distribution using the analytical model for the noise. 
The nearly perfect accuracy of the analytical noise model (see 
figure \ref{fignoise1}), which we have checked for
four cosmological models with different
smoothing scales and noise distribution, 
gives us confidence that the lensing signal can be 
extracted from forthcoming data --- either using deconvolution, or by 
comparing the forward convolution for a set of models. 

Analytical predictions of peak statistics would be valuable
in comparing theoretical predictions with observations. 
Reblinski et al (1999) have shown that
predictions of peak number densities based on the Press-Schechter
model agree with the simulations for the high-$\sigma$ peaks. 
Detailed analytical predictions of peak number densities and their
angular correlations by combining the Press-Schechter model and 
its extensions with our noise model would be useful. 
Further work is also needed to test the sensitivity of the
results to the shape of the dark matter power spectrum. 
The dependence on the redshift distribution of source galaxies  
needs to be computed as well --- since the level of
non-Gaussianity decreases for more distant galaxies, increasing the
redshift of source galaxies could mimic the effect of high-$\Omega$. 

To place the peaks approach in perspective, it is useful to 
compare it with the standard approach of measuring dark matter 
statistics using the entire field (without peak identification). The
peak statistics rely on only a subset of the available information 
(the location and height of peaks, and eventually their profile),
and obtaining cosmological
information from them requires additional theoretical modeling 
compared to field statistics. 
On the other hand, the use of peak statistics has both practical
and theoretical advantages. Peak statistics are likely to be robust
to certain kinds of systematics --- small, unknown errors in the galaxy
ellipticities that complicate the use of field
statistics. For example, in practice, the shear measured from 
ellipticity data is multiplied by a factor larger than unity to
account for the smearing by the point spread function. 
If this factor is estimated incorrectly,
it could change the height but probably not the location of the
peaks. It would then 
amount to a rescaling of the x-axis in the peak histograms, which does
not change the comparisons amongst the cosmological models.

On the theoretical side, peak statistics can
provide insights into the biasing of galaxies relative to the dark 
matter, by allowing us to consider the two distinct components of
biasing: first, the relation of galaxies to dark halos, 
and second, of halos to the dark matter. 
By combining the measured clustering of galaxies 
with that of dark halos measured through peak statistics from 
lensing data, the first step in the biasing of galaxies can 
be directly probed. For the second step
of relating halos to the dark matter, we will need to use successful 
measurements of field lensing, or to interpret the data using theoretical
models for the relation of halos to the dark matter. 

We have shown that the statistics of peaks provides a useful new
approach to wide field lensing. It is complementary 
to the standard statistics such as ellipticity correlations over
the field, and is directly linked to the projected distribution of 
dark matter halos. The characteristic non-Gaussian shape 
of the peak distribution (its asymmetric double peaked shape) 
makes it a powerful probe of the cosmological model as well as 
a useful test of the presence of systematics errors.

\acknowledgments

We are grateful to Uros Seljak, Peter Schneider, Ravi Sheth, 
Alex Szalay and Simon White for helpful discussions. 
We thank an anonymous referee for comments.

\end{document}